# REDUCING FAILURE PROBABILITY OF CLOUD STORAGE SERVICES USING MULTI-CLOUD

**Dissertation**

Submitted

in partial fulfillment

for the award of degree of

*Master of Technology*

*in Department of Computer Science and Engineering*

*(with specialization in Computer Science)*

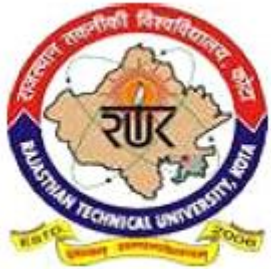

| Supervisor | Submitted By: |
|---|---|
| Mr. R.K. Banyal | Veena Rawat |
| Asst. Professor, CSE | Enrolment No.: 11E2UCCSF45P614 |

**Department of Computer Science and Engineering**

University College of Engineering, Kota

Rajasthan Technical University

June 2013

# Certificate

I hereby declare that the work, which is being presented in the Seminar, entitled "**Reducing Failure Probability of Cloud Storage Services using Multi-Cloud**" in partial fulfillment of "Master of Technology" with specialization in Computer Science and Engineering, submitted to the Department of Computer Science & Engineering, University College of Engineering, Rajasthan Technical University, Kota is a record of my own investigations carried under the guidance of Mr. R.K. Banyal, Assistant Professor, Computer Science, University College of Engineering, RTU Kota.

I have not submitted the matter presented in this Dissertation Review any where for the award of any other Degree.

Veena Rawat
Computer Science Engineering
Enrolment No.: 11E2UCCSF45P614
University College of Engineering,
RTU, Kota (Raj)

Supervisor:
R.K. Banyal
Assistant Professor, CSE
University College of Engineering,
RTU, Kota (Raj)

# Acknowledgment

It is my great privilege to express sincere gratitude & thanks to my supervisor **Mr. R.K. Banyal**, Assistant Professor, University College of Engineering, Rajasthan Technical University, Kota for his valuable guidance during each and every phase of this work. The keen observation and motivating style of critique and compliment kept me stimulated towards perfection. I thank him for the interest and energy that was committed to the seminar and for allowing me a wide academic freedom. I would also like to thank Mr. C.P. Gupta, Head, Department of Computer Science, UCE, RTU, Kota whose motivation and guidance to me in this work is unforgettable.

I would also thank my family for their help, support and patience. Not the least, I would thank Almighty for blessing me.

Veena Rawat

M.Tech IV Sem

Computer Science Engineering

Enrolment No.: 11E2UCCSF45P614

University College of Engineering,

RTU, Kota (Raj)

# Abstract

The work presents a model of a horizontal scaling in cloud storage and studies the optimal multi-cloud storage providers selection within a given budget. The model takes due consideration to many dimensions in cloud storage like cost, performance, security and privacy, and availability. While estimating the cost of cloud storage services, apart from storage cost, other costs like data in costs, data out costs, PUT and GET requests costs are also included.

Any information is valuable as long as it has related data. If related data are not put together, the information is meaningless as unrelated data has no value. The mapped information is required only by authenticated users. So there is no necessity to store related information together. If the relations of a database are fragmented into chunks and these chunks are stored at different cloud service providers, it could prevent from any privacy breach and the data stored will be secure. It would also reduce the data transfer costs as the entire data is not always required, for e.g. during updates. Also, instead of storage of chunks at a single CSP, if each chunk or fragment is stored at multiple CSPs it ensures availability and also permits concurrent access. Additionally, it would prevent financial loss during cloud outages and also prevent data lock-in. Replicating data chunks at multiple clouds situated at geographically different locations would also have an additional decrease in response time.

The work attempts to select multiple cloud service providers within a given budget so as to ensure maximum availability of data. The entire data can be stored at each of the data centers selected depending on the budget when there is no security or privacy issue. Data can also be stored in chunks by replicating each data chunk at two or more cloud service providers. Different chunks can be replicated at different service providers. The work also attempts to select various cloud service providers to ensure maximum valid data chunks within a given budget.

# Table of Contents

# List of Figures

# List of Tables

# Chapter 1

# INTRODUCTION

## 1.1 Need for Multi-Cloud Storage

Many applications today such as Web applications, Mobile applications are not limited to any geographical boundary. The customers of these applications may be at far flung places. In such cases, if one wants to expand globally, then Multi Cloud strategies are a boon. While this enables us to reach distributed markets, new challenges related to latency, performance, pricing, availability crop up. Every customer residing at far flung places needs good performance not only for certain period of time, but a consistent good performance is required.

Ensuring consistent performance and high availability are the two major challenges for global expansion. A single cloud service provider or a single delivery network cannot be trusted for this task. Delivering maximum performance globally round the clock is not possible even by most resilient cloud service provider. The only solution is adopting multi-clouds which would distribute the performance and availability threats across global public and private data centers. Global performance is not a representation of a single cloud service providers but a judicious assimilation of many.

In order to drive revenue and other benefits that are very closely related with performance and availability, a multi-cloud strategy is essential. When options are available, performance and availability differences are in dollars.

One cannot rely on one cloud service provider to fulfill the requirements of customers belonging to different geographic regions. So, what is required is an assimilated network of multiple clouds.

Recent high-profile cloud outages are unmitigated manifestation of the need for multi-sourcing, although many companies are still relying on single-source providers. It's an avoidable threat.

A multi-cloud strategy permits one to manager traffic across data centers, clouds and delivery networks to manage costs and optimize price-to-performance ratios.

Web performance means everything when revenue and brands are on the line. And for companies operating globally, performance is best enabled with a multi-cloud strategy. Better web performance can be enabled through multi-cloud strategies that span data centers, delivery networks and cloud providers.

IT administrators are often concerned about losing authority of data placed in the cloud. Whether data will be easily available, mainly in the event of a cloud provider outage? What if that outage prevails for prolonged period of time? Employing redundant, independent systems can help a lot in order to reduce the threat. Users who use public clouds for storage can easily attain this by using services of two or more different cloud providers.

An unhappy thing about life is that things collapse. This is true for Cloud Computing also: no matter how much better uptime or availability or performance is offered by cloud providers, these services collapse eventually. Preparedness is the only thing we can do. Building redundancy in cloud based application is a part of preparedness, but in clouds this redundancy is limited to running several redundant copies on separate data centers of the same cloud provider. Having multiple data centers at different geographical locations by large cloud service providers is one approach to the solution of this problem. Another possible solution could be adoption of multi-cloud strategy.

By using services of multiple cloud service providers, redundancy is achieved at a new abstraction level. In order to host our cloud servers, if we select data centers from different providers, we can effectively do away with the threats related with business continuity of the cloud service provider, threats concerning electricity suppliers, "data center" managerial issues, networking providers. This is possible as each service provider works independently.

Other threats which are correlated with a single cloud service provider are also reduced. Cloud works on virtualization. If in case, any vulnerability is detected on our infrastructure provider, and a multi-cloud strategy is adopted, one can immediately switch on to the other provider without any impact to the operations. The same thing can happen in the case of price hikes or changes in the terms and conditions of our current service provider. Stopping the services of the current service provider and switching to the service of the other provider is a good solution.

Adoption of multi-cloud strategy was difficult 3-4 years ago because the cloud providers operated on closed architectures that were proprietary. Migration from one cloud service provider to another was too difficult. Cloud providers worked on different platforms and interoperability among them was not possible. The only solution was to download everything from the current service provider, build new virtual machine at another provider and then upload the entire data again. Today with the advent of interoperability, these barriers are no more.

Now data can be migrated from one service provider to another very easily. The data stored in one service provider's virtual machine can be copied into another service provider's virtual machine very easily. Even, the facilities for uploading and downloading the entire virtual machines is provided by the cloud service provider.

Cloud providers focus on delivering "3 Nines". This availability alone is not enough to meet SLAs of enterprise customers. As shown in Table 1.1 high end applications require "6 Nines" availability and security in terms of data privacy, and performance. Business operations such as 24X7 online retailers, email applications cannot withstand this level of downtime.

*Table 1.1: Number of Nines and Downtime in Seconds*

| Nines | Percentage | Downtime in a year |
|---|---|---|
| 2 | 99% | 3.65 days |
| 3 | 99.9% | 8.75 hours |
| 4 | 99.99% | 52 minutes |
| 5 | 99.999% | 5 minutes |
| 6 | 99.9999% | 31 seconds |

Although working with multi-cloud service providers seems to be beneficial, some logistical problems may creep in. Open source tools or vendor-agnostic tools and cloud management services can do a lot in resolving these problems.

Another problem is that different cloud providers offer different types of services. Also additional tools and services are provided by some of the service providers. So, if we think of working with multiple cloud service providers, we may have to resort to the least services offered by them. But, such problems can be easily solved by using additional vendor-agnostic software applications which can easily be used by all of our cloud service providers. The major hazards at the forefront of IT concerns are data lock-in and cloud outages which can easily be handled by multi-clouds.

The rest of this dissertation report is organized as follows. Chapter 2 gives a description of the literature survey. Chapter 3 describes the methodology used to fragment data into chunks using vertical fragmentation using privacy constraints and horizontal fragmentation. Chapter 4 describes the dynamic programming based mathematical formulation of the two problems and their algorithms. A detailed description and experimental results of two algorithms is given in Chapter 5. Finally in Chapter 6, the conclusions drawn from the experiments are discussed.

## 1.2 Objectives

Reduce Failure Probability of cloud storage services using multi-clouds.

i) Reduce Failure probability of data within fixed budget through selection of multiple cloud service providers.

ii) Ensure privacy of user data on the cloud through fragmentation of data into chunks before replicating them on the cloud service providers.

iii) Maximise expected value of data chunks by replicating data chunks among cloud service providers within fixed budget.

## 1.3 Scope of work

The proposed work is limited to ensuring privacy of user data stored in the cloud by fragmentation of data horizontally and vertically using privacy constraints and distributing the data among multi-clouds such that none of the CSP has full amount of data. So the data stored with any one CSP is of no value to him. The work assists in selecting the best service providers in terms of response time, availability and cost within a given budget. The work is limited to one region only. The same can be repeated for different regions around the globe as different cloud service providers provide different response time at different places. Also the data about QoS specified in terms of response time has been taken up directly as mentioned by various services who compare various cloud service providers.

The rest of this dissertation report is organized as follows. Chapter 2 gives a description of the literature survey. Chapter 3 describes the methodology used to fragment data into chunks using vertical fragmentation using privacy constraints and horizontal fragmentation. Chapter 4 describes the dynamic programming based mathematical formulation of the two problems and their algorithms. A detailed description and experimental results of two algorithms is given in Chapter 5. Finally in Chapter 6, the conclusions drawn from the experiments are discussed.

# Chapter 2

# LITERATURE SURVEY

Hussam Abu-Libdeh et. al in their work "RACS" use RAID-like techniques used by disks and file systems, but at the cloud storage level. They argue that striping user data across multiple providers can allow customers to avoid vendor lock-in, reduce the cost of switching providers, and better tolerate provider outages or failures [1].

Thanasis G. Papaioannou, Nicolas Bonvin and Karl Aberer introduce "Scalia", a cloud storage brokerage solution that continuously adapts the placement of data based on its access pattern and subject to optimization objectives, such as storage costs. Scalia efficiently considers repositioning of only selected objects that may significantly lower the storage cost. By extensive simulation experiments, they prove the cost-effectiveness of Scalia against static placements and its proximity to the ideal data placement in various scenarios of data access patterns, of available cloud storage solutions and of failures [2].

Lluis Pamies-Juarez, Pedro Garcia-Lopez, Marc Sanchez-Artigas, Blas Herrera in their work "Towards the Design of Optimal Data Redundancy Schemes for Heterogeneous Cloud Storage Infrastructures" analyze how distributed redundancy schemes can be optimally deployed over heterogeneous infrastructures. Specifically, they are interested in infrastructures where nodes present different online availabilities. Considering these heterogeneities, they present a mechanism to measure data availability more precisely than existing works. Using this mechanism, they infer the optimal data placement policy that reduces the redundancy used, and then its associated overheads up to 70% [3].

Stefan Wind, Klaus Turowski and Jonas Repschläger, Rüdiger Zarnekow in their work "Target Dimensions of Cloud Computing" developed target dimensions for cloud computing, based on an international literature analysis and interviews with experts. In special, they have been explained using Infrastructure as a Service. These dimensions help enterprises to become clear about their requirements on cloud computing and do further steps, like classifying appropriate providers [4].

Zia ur Rehman and Omar K. Hussain, Farookh K. Hussain in their work "Iaas Cloud Selection using MCDM Methods" use multi-criteria decision-making methods for IaaS cloud service selection in a case study which contains five basic performance measurements of thirteen cloud services by a third party monitoring service. They demonstrate the use of these multi-criteria methods for cloud service selection and compare the results obtained by using each method to find out how the choice of a particular MCDM method affects the outcome of the decision-making process for IaaS cloud service selection [5].

Kevin D. Bowers, Ari Juels and Alina Oprea in their work "HAIL" (High-Availability and Integrity Layer), a distributed cryptographic system that allows a set of servers to prove to a client that a stored file is intact and retrievable. HAIL strengthens, formally unifies, and streamlines distinct approaches from the cryptographic and distributed-systems communities. They show how HAIL improves on the security and efficiency of existing tools, like Proofs of Retrievability (PORs) deployed on individual servers [6].

Chia-Wei Chang, Pangfeng Liu, Jan-Jan Wu, "Probability-Based Cloud Storage Providers Selection Algorithms with Maximum Availability," select cloud service providers based on cost and availability metrics.[7]

Carlo Curino, Evan Jones, Yang Zhang, Eugene Wu in "Relational Cloud: The Case for a Database Service" in order to allow workloads to scale across multiple computing nodes, divide data into partitions that maximize transaction/query performance. They have developed a new graph-based data partitioning algorithm for transaction-oriented workloads that groups data items according to their frequency of co-access within transactions/queries [8].

G. Aggarwal, M. Bawa, P. Ganesan, H. Garcia-Molina, K. Kenthapadi, R. Motwani, U. Srivastava, D. Thomas, Y. Xu in "Two Can Keep a Secret: A Distributed Architecture for Secure Database Services" perform efficient partitioning of data using privacy constraints on distributed database[9].

Subashini, S. and V. Kavitha in "A Metadata Based Storage Model for Securing Data in Cloud Environment" in order to eliminate the disadvantage of storing all data of a client to the same provider, split data into chunks and distribute them among multiple cloud providers [10].

Ms. P. R. Bhuyar, Dr. A.D. Gawande, Prof. A.B.Deshmukh in "Horizontal Fragmentation Techniques in Distributed Database" fragment a relation horizontally according to locality of precedence of its attributes [11].

Himel Dev, Tanmoy Sen, Madhusudan Basak and Mohammed Eunus Ali in "An Approach to Protect the Privacy of Cloud Data from Data Mining Based Attacks" inside the Cloud Data Distributor provide each chunk a unique virtual id and this id is used to identify the chunk within the Cloud Data Distributor and Cloud Providers. This virtualization conceals the identity of a client from the provider [12].

## 2.1 Cloud Storage Metrics

Saurabh Kumar Garg, Steev Versteeg, Rajkumar Buyya in their work "A framework for ranking of Cloud computing services" provide Cloud measurement metrics [13].

*The Cloud Service Measurement Index*

This framework provides a holistic view of QoS needed by the customers for selecting a Cloud service provider based on: Accountability, Agility, Assurance of Service, Cost, Performance, Security and Privacy, and Usability.

*Accountability:* Accountability *is* used for measuring and scoring services, that include auditability compliance, data ownership, provider ethicality, sustainability etc.

*Agility:* Agility in SMI is measured as a rate of change metric showing how quickly new capabilities are integrated into IT as needed by the business.

*Cost:* Cost is clearly one of the vital attributes for IT and the business. Cost tends to be the single most quantifiable metric today, but it is important to express cost in the characteristics which are relevant to a particular business organization.

*Performance:* There are many different solutions offered by Cloud providers addressing the IT needs of different organizations. Each solution has different performance in terms of functionality, service response time and accuracy. Organizations need to understand how their applications will perform on the different Clouds and whether these deployments meet their expectations.

*Assurance:* This characteristic indicates the likelihood of a Cloud service performing as expected or promised in the SLA. Every organization looks to expand their business and provide better services to their customers. Therefore, reliability, resiliency and service stability are important factors in selecting Cloud services.

*Security and Privacy:* Data protection and privacy are important concerns for nearly every organization. Hosting data under another organization's control is always a critical issue which requires stringent security policies employed by Cloud providers. For instance, financial organizations generally require compliance with regulations involving data integrity and privacy. Security and Privacy is multi-dimensional in nature and includes many attributes such as protecting confidentiality and privacy, data integrity and availability.

*Usability:* For the rapid adoption of Cloud services, the usability plays an important role. The easier to use and learn a Cloud service is, the faster an organization can switch to it. The usability of a Cloud service can depend on multiple factors such as Accessibility, Installability, Learnability, and Operatibility.

## 2.2 Cloud Comparison Metrics by CloudCmp

While many public cloud providers offer pay-as-you-go computing, their varying approaches to infrastructure, virtualization, and software services lead to a problem of plenty. To help customers pick a cloud that fits their needs, Ang Li et.al develop CloudCmp, a systematic comparator of the performance and cost of cloud providers [14].

CloudCmp measures the elastic computing, persistent storage, and networking services offered by a cloud along metrics that directly reflect their impact on the performance of customer applications. CloudCmp strives to ensure fairness, representativeness, and compliance of these measurements while limiting measurement cost.

CloudCmp can guide customers in selecting the best-performing provider for their applications. They use three metrics to compare the performance and cost of storage services: operation response time, time to consistency, and cost per operation.

*Operation response time*: This metrics measures how long it takes for a storage operation to finish. They measure operations that are commonly supported by providers and are popular with customers. They include basic read and write operations for each storage service. For table storage service they also use an SQL-style query to test the performance of table lookup. These operations account for over 90% of the storage operations used by a realistic e-commerce application.

*Time to consistency:* This metrics measures the time between when a datum is written to the storage service and when all reads for the datum return consistent and valid results. Such information is useful to cloud customers, because their applications may require data to be immediately available with a strong consistency guarantee. Except for AppEngine, cloud providers do not support storage services that span multiple data centers. Therefore, they focus on consistency when the reads and writes are both done from instances inside the same data center.

*Cost per operation*: The final metrics Cost per Operation measures how much each storage operation costs. With this metrics, a customer can compare the cost-effectiveness across providers.

## 2.3 Cloud Comparison Metrics by Nasuni - An Enterprise Storage Provider

Cloud storage offers a unique and advanced set of benefits, including infinite scalability and global access. Still, it is also a relatively new technology in a market that is rapidly evolving. In order to take advantage of such technology and ensure quality, Nasuni engineers conduct frequent, ongoing testing and monitoring of Cloud Storage Providers.

Nasuni, a provider of enterprise storage to large, distributed organizations partners with Cloud Storage Providers to achieve the best possible product at the most cost-effective price [15]. Organizations considering cloud storage as part of their storage infrastructure should consider these same tradeoffs when comparing CSPs. Nasuni evaluates five key components of each CSP's offering: Functionality, Price, Performance, Availability and Scalability

*Functionality*

While most interactions that an enterprise has with CSPs consist of simple API commands (GET, PUT and DELETE), an organization should consider a broader range of functionality when comparing cloud storage providers. Many companies today are global operations with offices around the world in a wide variety of localities, from major metropolitan areas to remote villages. To serve such users, cloud service providers need to maintain access points around the world and support meaningful cross-geography replication. Two copies of a file in a single datacenter is not geographic redundancy. In addition, organizations that expect to make meaningful use of cloud storage in their environment should also evaluate features of potential providers such as their API-based account creation and account management processes, availability of libraries and software to access data, the sophistication of their billing schemes and other aspects that help operations teams to ensure a smooth experience for their users and applications.

*Price*

Cloud storage architecture is fundamentally different from traditional storage; consequently, it is also priced differently from conventional storage. Instead of charging price per raw TB (as with traditional storage hardware), most CSPs charge based on GB stored per month. However, pricing is typically more complicated than a simple count of GB per month, often adding compute costs (to process API commands) and network costs (to move data to and from the cloud storage). While this pricing model is cost effective because it charges customers only for the resources that they use, it makes predicting future costs a complex endeavor due to the variability of applications and use-cases. Although some vendors provide tools to help estimate costs, every customer's use-case is unique, so one-size-fits-all tools provide poor predictions. Unless the organization is working with a provider that offers a simplified pricing scheme, it is best to conduct initial tests with a minimal investment and then extrapolate from those results to develop a more accurate pricing estimation model. Price itself is a very small part of a CSP comparison and may be the last part of a decision. Commodity offerings combined with competitive activity are driving costs down rapidly, however functionality and performance still vary significantly. When evaluating a CSP, price is easy to change and negotiate – functionality and performance are not.

*Performance*

Performance is the primary yardstick by which Nasuni measures any publicly available CSP, testing the operation and stability of CSPs over long periods of time. In fact, Nasuni has been testing and comparing CSPs since 2009. Before considering any CSP for use in a production environment, it must meet minimum performance benchmarks across three areas:

Write/Read/Delete Benchmark: This simple test measures the raw ability of each CSP to handle thousands of writes, reads and deletes (W/R/D). We test each CSP with files of varying sizes ,1 KB, 10 KB ,100 KB ,1 MB ,10 MB ,100 MB ,1 GB using varying levels of concurrency: 1 Thread ,10 Threads ,25 Threads , 50 Threads The Write/Read/Delete benchmark test runs for twelve hours, using multiple testing machine instances and several non-serial test runs to reduce the likelihood that external network issues could bias the results.

*Availability*

This test takes place over a 30-day period and measures each CSP's response time to a single W/R/D process at 60-second intervals:

- Write a randomly generated 1 KB file
- Read a randomly selected previously written file
- Delete a selected file

Reading and deleting a random file forces each CSP to prove their ability to be responsive to all of the data, all of the time, and not merely to the last piece of cached data. This test calculates the entire time required to complete the three requests, including any required retries. This ensures examination of not only responsiveness but also of CSP reliability and latency.

*Scalability* Similar to the availability test, this is also an extended test that measures each CSP's ability to perform consistently as the number of objects under management increases. Performance under increasing object counts is often the Achilles heel of a cloud storage system, and this test measures each CSP's ability to maintain performance levels as the total number of objects stored in a single container increases to hundreds of millions.

## 2.4 Cloud Comparison by Cedexis- A Radar Community

The Cedexis Radar community measures many cloud, content delivery, and private platforms, as users experience them, from over 32,000 networks around the globe [16]. The result is a real-time picture of global performance for all of these platforms from nearly every network in every country in the world. Performance in terms of response time measured in milliseconds in different countries are shown below in Table 2.1.

*Table 2.1: Response time in milliseconds as measured on 11 Apr, 2013*

| Cloud Service Provider | China | Brazil | Canada | US | Russia | Australia | Mexico | India | Algeria | South Africa |
|---|---|---|---|---|---|---|---|---|---|---|
| Windows Azure | 132 | 323 | 112 | 153 | 158 | 337 | 187 | 276 | 365 | 429 |
| InterNap AgileCloud | 182 | 391 | 170 | 212 | 151 | 394 | 302 | 252 | 281 | 417 |
| Savvis | 176 | 311 | 94 | 141 | 153 | 317 | 183 | 467 | 388 | 421 |
| Amazon | 266 | 153 | 92 | 140 | 162 | 135 | 181 | 282 | 361 | 442 |
| SoftLayer | 311 | 298 | 90 | 136 | 144 | 327 | 169 | 400 | 376 | 400 |
| PhoenixNAP | 320 | 309 | 94 | 143 | 155 | 334 | 178 | 425 | 321 | 448 |
| Rackspace | 393 | 333 | 88 | 149 | 171 | 364 | 169 | 421 | 345 | 433 |
| Joyent | 373 | 299 | 92 | 143 | 142 | 312 | 170 | 421 | 312 | 402 |
| Cloud Sigma | 469 | 351 | 149 | 227 | 159 | 332 | 175 | 518 | 317 | 466 |
| Google App Engine | 1298 | 1150 | 880 | 1044 | 278 | 1214 | 1004 | 1261 | 455 | 536 |
| ProfitBricks | 387 | 350 | 141 | 168 | 150 | 317 | 188 | 485 | 350 | 403 |

From the above data collected from the official website, we see variations in the response time of various CSP. It means that the cost of storage, availability and response time may vary over time as shown in Table 2.2. So the data regarding these metrics must be updated at regular intervals in order to get the most latest information and choose the best service providers.

In order to reduce the response time, we select the CSPs who provide less response time and then use this data further to find the best storage availability at minimum cost.

*Table 2.2: Response time in milliseconds as measured on 18 July, 2013*

| Cloud Service Provider 18-07-2013 | China | Brazil | Canada | US | Russia | Australia | Mexico | India | Algeria | South Africa |
|---|---|---|---|---|---|---|---|---|---|---|
| Windows Azure | 155 | 297 | 85 | 154 | 181 | 246 | 212 | 253 | 234 | 334 |
| InterNap AgileCloud | 333 | 268 | 81 | 140 | 163 | 309 | 173 | 262 | 285 | 420 |
| Savvis | 209 | 289 | 77 | 140 | 172 | 271 | 189 | 391 | 235 | 414 |
| Amazon | 274 | 148 | 74 | 143 | 169 | 125 | 194 | 247 | 289 | 485 |
| SoftLayer | 304 | 282 | 74 | 139 | 153 | 253 | 183 | 284 | 223 | 378 |
| PhoenixNAP | 433 | 287 | 74 | 141 | 153 | 292 | 187 | 348 | 264 | 398 |
| Rackspace | 368 | 298 | 74 | 140 | 166 | 347 | 202 | 344 | 264 | 444 |
| Joyent | 303 | 264 | 78 | 140 | 155 | 289 | 185 | 338 | 172 | 431 |
| Cloud Sigma | 367 | 340 | 162 | 170 | 162 | 304 | 192 | 340 | 298 | 405 |
| Google App Engine | 1153 | 1031 | 630 | 772 | 278 | 1112 | 942 | 1249 | 443 | 530 |
| ProfitBricks | 348 | 332 | 124 | 183 | 158 | 293 | 191 | 349 | 258 | 465 |

## 2.5 Security

Security has always been a major focus for CSPs. The various types of certifications regarding security compliance [17][18][19][20][21][22][23][24] are given in Table 2.3.

*Table 2.3: Cloud Service Providers and their security Certifications*

| Cloud Service Provider | Certification |
|---|---|
| Amazon | SSAE 16 ,ISO 27001 certification |
| Google App engine | SAS 70, SSAE 16, and ISAE 3402 compliant. |
| GoGrid | SSAE 16 Type II certified |
| Rackspace | SSAE 16 Type II SOC 1 |
| HP Cloud | SAS 70 Type II |
| Microsoft Azure | SSAE 16 / ISAE 3402 ,ISO/IEC 27001:2005, |
| InterNap AgileCloud | SSAE 16 Type II |
| Joyent | SSAE 16, PCI DSS LEVEL 1 |
| PhoenixNAP | SSAE 16, PCI DSS |
| savvisdirect | SSAE 16 Type II SOC 1 |

The SAS 70 certification acknowledges that the policies and control measures in place, sufficiently meet operational standards. Most importantly, an SAS 70 Type II audit affirms to potential clients that the provider is qualified to handle enterprise-class hosting chores, is serious about its service commitment, and is willing to subject itself to an extensive and unpleasant audit for the benefit of its clients.

Statement on Standards for Attestation Engagements (**SSAE 16)** is an enhancement to the current standard for Reporting on Controls at a Service Organization, the SAS70. The changes made to the standard will bring company, and the rest of the companies in the US, up to date with new international service organization reporting standards, the ISAE 3402. The adjustments made from SAS 70 to SSAE 16 will help to compete on an international level; allowing companies around the world to give you their business with complete confidence. SSAE16 is now effective as of June 15, 2011.

ISAE 3402 is a standard put forth by the International Auditing and Assurance Standards Board (IAASB), a standard-setting board within the International Federation of Accountants (IFAC). ISAE 3402 is an excellent example of the continuing migration in adopting true global accounting standards. ISAE 3402: The International Standard on Assurance Engagements, is to be the new globally recognized standard for assurance reporting on service organizations.

An ISAE 3402 Type 1 Report is known as the Report on the description and design of controls at a service organization.

ISO/IEC 27001 is the formal set of specifications against which organizations may seek independent certification of their Information Security Management System (ISMS).

# Chapter 3

# DATA FRAGMENTATION INTO CHUNKS

## 3.1 Data Storage Unit

*File as a unit of storage :* Replications of the entire file at multiple CSP's is beneficial if the file does not contain sensitive data and the queries require all the data. If the file is stored at only one CSP and is not replicated at more than one CSP, a single CSP will get a high volume of remote data accesses. Storing at multiple CSP ensures the availability of data as well as permits concurrent access.

*Chunks of file as a unit of storage:* Users require only a subset or a fragment of a file and the locality of access is defined on those fragments. Chunk storage permits a number of users to execute concurrently since the users will access different portions of a file. Parallel execution of a single query is also possible. Fragments of a file is usually the appropriate unit of storage. They aim to improve security, reliability, storage costs, update costs and communication costs.

If the data in the file falls under the category 'Normal' and the queries do not require all the data, chunk storage of data can be considered, where the data can be fragmented depending on the type or queries to be executed be it Horizontal or Vertical or Hybrid fragmentation. But if the data to be stored is 'Sensitive', then simple Horizontal, Vertical or Hybrid fragmentation would not provide the required security of data.

## 3.2 Data Storage Model

Consider the data is stored at a single Cloud Database as a Service provider. Then there is a single point of failure which will affect data availability. Availability is also an important issue if he runs out of business. Cloud service customers cannot rely on single CSP to ensure storage of vital data. If the database is stored at two DBaaS providers, there are chances that the two CSPs can act together secretly to achieve a fraudulent purpose and exchange the part of the data with each other and reconstruct the whole data.

In our approach, the client does not have to trust the administrators of any cloud service providers to guarantee privacy. So long as an adversary does not gain access to all the data, data privacy is fully protected. If the client were to obtain database services from different vendors, the chances of an adversary breaking into all the service providers, is greatly reduced. Furthermore, the insider attacks at any one of the cloud service providers do not compromise the security of the system as a whole.

If database security is taken care of by the customer, it also helps the cloud service provider by limiting their liability in case of break-ins into their system. If the service provider is not able to find any valuable information from the contents of the database, nor will the outsider be. Existing proposals for secure database service are based on encryption. Although, these attempts are good at securing data in the cloud, they cause large overheads in query processing. Weak encryption algorithms that allow efficient queries leak far too much information and thus do not preserve privacy. On the other hand strong encryption algorithms often necessitate resorting to Plan A for queries, fetching the entire database from the servers which is simply too expensive. Despite increasing processor speeds encryption and decryption are not exactly cheap. New approach is to allow the client to partition its data across three or more logically independent cloud storage systems.

### 3.3 Data Privacy

Each file has a privacy level: 'Normal', 'Sensitive' or 'Critical'. The data which has low value to cloud service providers or attackers and can be allowed to be stored as public data is considered as 'Normal'. The data which is having high value is considered as 'Critical' and the data which has value when mapped with other data is considered as 'Sensitive'. The data which maps 'Sensitive' or 'Critical' data to 'Normal' data is also considered as 'Sensitive'.

The steps to ensure data privacy consists of Categorization, Fragmentation, Distribution, and Replication. Categorize user data as 'Normal', 'Sensitive' or 'Critical'. Split user data into chunks based on the categorization and provide these chunks to CSPs providing Database as a Service. Fragmentation of data is performed in such a fashion so as to ensure that the exposure of the contents of anyone database does not result in a violation of privacy. The presence of three or more cloud service providers enable efficient semantic attribute decomposition, or attribute encoding of sensitive attributes. For example, we can store telephone number by segregating area code at one CSP and telephone number at another CSP. The presence of multiple cloud service providers also enable the storage of many attribute values in unencrypted form. Typically the exposure of a set of attribute values corresponding to a tuple may result in privacy violation while the exposure of only some subsets of it may be harmless. For example individual's name and his credit card number may be a serious privacy violation. However, exposing the name alone or the credit card number alone may not be a big deal. In such cases we may place individual's name in one CSP while storing his credit card number in another avoiding having to encrypt either attribute. A consequence is that queries involving both names and credit card number may be executed far more efficiently than if the attributes had been encrypted.

Distribution is done according to the sensitivity of data and the reliability of CSP. Reliability is defined in terms of reputation and reliability of the CSP. Distribution restricts an attacker from having access to sufficient number of chunks of data and thus prevents successful extraction of valuable information.

### 3.4 Architecture

Architecture as shown in Figure 3.1 consists of trusted client as well as three or more cloud service providers that provide Database as a Service. The Database as a Service providers provide reliable content storage and data management but are not trusted by the clients to preserve content privacy. The client does not store any persistent data but stores a mapping table describing the storage of various fragments location, their names etc. However the client has access to cheap hardware providing processing power as well as temporary storage and functionality in terms of offering a DBMS frontend, reformulating and optimizing queries and post processing query results, all of which are fairly cheap and can be performed using inexpensive hardware. The client executes queries by transmitting appropriate sub queries to each database and then piecing together the result obtained from the Cloud service providers at the client side.

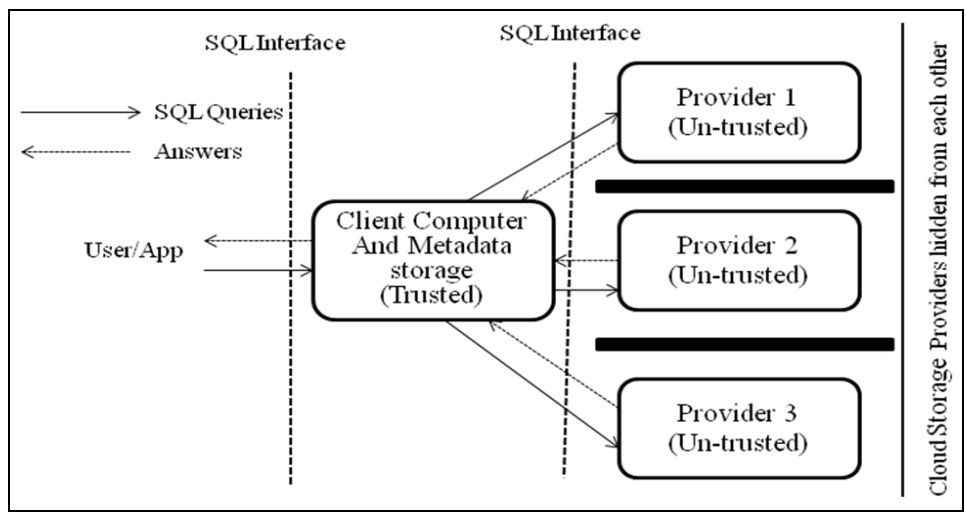

*Figure 3.1: Multi Cloud Architecture*

### 3.5 Relational Decomposition

Data fragmentation and distribution among multiple CSPs is performed for ensuring security and availability of data in cloud. There are different techniques to partition a relation $R = (A_1, A_2, A_3, ... A_n)$. Traditional decomposition methods are Horizontal, Vertical and Hybrid fragmentation. Horizontal fragmentation partitions a relation along tuples. Vertical fragmentation partitions a relation along attributes and a Mixed/Hybrid fragmentation is a combination of Horizontal and Vertical fragmentations. The fragments should be constructed such that they fulfill

Completeness, Reconstruction and Disjointness properties.

Horizontal fragmentation is done based on the selection conditions in the queries to reduce the amount of data during transfers. Horizontal fragmentation has limited use in enabling privacy preserving but it can be of great use in reducing communication costs. Whenever Reads or Writes or Delete operations are performed, they are not always on the entire relation. Horizontal fragmentation is done according to the workload behavior of the queries.

Vertical fragmentation requires key attributes to be present in the sub relations. Key attributes may themselves be sensitive information. A single attribute may become a privacy constraint. In that case, it cannot be stored in open. It can be stored either by encoding the attribute or by storing the hash of the attribute, or by performing semantic attribute decomposition. If it is a primary key attribute, in such a case, introduce a unique tuple ID. We can generate random numbers as tupleIDs ensuring that tupleIDs not already exist. Vertical fragmentation may require semantic attribute decomposition where an attribute $A$ is split into two attributes $A1$ and $A2$. Attribute $A1$ is stored in one of the sub relation and attribute $A2$ is stored in another. For example while storing credit card number issuer identification number is stored in one of the sub relation and individual account identifier and check digit is stored other. Semantic Attribute decomposition will also benefit selection queries based on individual account number or queries that perform aggregation when grouping by issuer identification number could benefit greatly from $A2$ attribute. In absence of $A2$, if credit card numbers were encrypted, query processing becomes more expensive. Attribute encoding can also be used for attributes that need to be kept private. For e.g. salary. Encode salary $s$ as $s_1$ and $s_2$ where $s_1 = s + r$ and $s_2 = r$. Store $s_1$ and $s_2$ in separate sub relations.

# Chapter 4

# METHODOLOGY

## 4.1 Specifying the Privacy Constraints

Privacy requirements on a relational schema $R$ are specified as a set or privacy constraints. Each privacy constraint is listed as a set of attributes which alone or together may have some value. The decomposition of relation $R$ should be such that for each privacy constraint on a relational schema $R$, all the attributes of a privacy constraint should not be a part of any sub relational schema. Some of the attributes of a constraint may be open, some may be encoded and some may be semantically decomposed but all the attributes of a privacy constraint cannot be together in any sub relation schema.

Consider a database in a bank consisting of user information along side with the credit card information.

- A Customer Table {CustomerId (Primary Key (PK)), CustomerName, CustomerAddress, CustomerPhone, CustomerDOB}

- A Membership Table {CustomerId (Primary & Foreign Key (FK)), Pwd, PwdQuestion, PwdAnswer}

Identify the privacy constraints on each table and then perform vertical fragmentation.

*1) Customer table Constraints*

a)   {CustomerPhone} is a sensitive information.

b)   {CustomerName and CustomerAddress}, {CustomerName and CustomerDOB}

c)   {CustomerAddress, CustomerPhone, Customer DOB}

CustomerPhone is a single privacy constraint and cannot be stored in clear. So it can be stored by semantically decomposing it into Area Code and Telephone number. The constraints specified in (b) and (c) can be addressed by vertical fragmentation of attributes. $R1$( CustomerId, CustomerName), $R2$( Customer Id, CustomerAddress, CustomerTelephoneAreaCode), $R3$(CustomerId, Customer TelephoneNo, CustomerDOB).

*2) Membership table Constraints*

a)   {Pwd} is a sensitive attribute.

b)   {PwdQuestion, PwdAnswer}

This table alone has no importance. But if the two cloud service providers collude, it has juicy information. So classify it as sensitive. Password is a sensitive attribute and cannot be stored in open. So store the hash value of the password. The constraints specified in (b) can be addressed by vertical fragmentation of attributes. $R1$ (CustomerId, Pwd#, PwdQuestion), $R2$ (CustomerId, Pwd# ,PwdAnswer).

*Distribution and Replication:* Cloud providers focus on delivering "3 Nines". This availability alone is not enough to meet SLAs of enterprise customers. High end applications require "4 Nines" availability. In order to ensure this high availability, after decomposition the client reformulates the queries and then replicates each decomposed relation (chunk) to at least two CSPs.

Replication of each chunk is done at more than one cloud service provider so as to increase cloud availability from 3 nines i.e. 99.9 % to at least 4 nines i.e. 99.99 %.

If one of the chunk storage providers goes down, the other chunk storage provider will provide the data chunks that were stored on the failed server. The client also maintains a mapping table of the various relations, chunks names, sequence of chunks and storage locations. Each chunk is given a random name. So even if the CSPs collude with each other and exchange the part of the data with each other, they cannot reconstruct the whole data. Even if the adversary is able to find out some information from all the chunks, he is not aware of the proper order of the chunks in making the information have some value. So the data will be secure. Splitting data into smaller chunks restricts data mining attacks also to a great extent as they contain insufficient amount of data.

## 4.2 Dynamic Programming

Dynamic programming is used to find solution to problems by breaking them into smaller sub-problems and solving those sub-problems. The solution to any sub problem is not computed more than once. Instead, the computed solutions are saved in a table so that they can be reused. Dynamic programming works well when the sub-problems are not independent [24] . Dynamic Programming is typically applied to optimization problems, where the goal is to find a solution among many possible candidates that minimizes or maximizes some particular value. Such solutions are said to be optimal. There may be more than one optimal solution. Dynamic programming works by solving sub problems and using the results of those sub problems to more quickly calculate the solution to a larger problem. Unlike the divide-and-conquer paradigm (which also uses the idea of solving sub problems), dynamic programming typically involves solving all possible sub problems rather than a small portion. Often, dynamic programming algorithms are visualized as "filling an array" where each element of the array is the result of a sub problem that can later be reused rather than recalculated.

## 4.3 QoS Criteria

The various QoS criteria can be viewed in a multidimensional space. The filtration process on each dimension can then be carried out which reduces the number of Cloud service providers satisfying the QoS criteria required by the client. The filtration process continues until the two dimensional data i.e. availability and cost is arrived. This dataset of Cloud Service providers meets all the QoS constraints as specified by the client.

## 4.4 Failure Probability and Cost of Storage of the popular CSPs.

Based on the information provided by various CSP, we summarize the failure probability and the cost of the popular CSPs.

From the data provided by various CSPs[26][27][28][29][30][31][32][33][34][35], we calculate the storage cost for storing the same amount of data, data transfer out, GET and PUT requests per month as mentioned below and present in a summarized form as shown in

- Data to be stored : 51.66 TB per month
- Data transfer out: 2 TB per day for 31 days
- Put request :1000 files each day for 31 days
- Get request: 20000 files each day for 31 days

| Cloud Service Provider | Availability | Failure Probability | -log (Failure Probability) | Cost of storage (51.66 TB per month) | Cost per TB in dollars |
|---|---|---|---|---|---|
| Amazon | 99.99 | 0.01 | 2 | 13919 | 269 |
| Google App Engine | 99.9 | 0.1 | 1 | 13511 | 262 |
| Gogrid | 99.99 | 0.01 | 2 | 12150 | 235 |
| Rackspace | 99.9 | 0.1 | 1 | 12381 | 240 |
| HP | 99.95 | 0.05 | 1.30103 | 12544 | 243 |
| Windows Azure | 99.9 | 0.1 | 1 | 10353 | 200 |
| Savvis | 99.95 | 0.05 | 1.30103 | 12648 | 245 |
| Internap | 99.95 | 0.05 | 1.30103 | 11796 | 229 |
| Softlayer | 99.9 | 0.1 | 1 | 11643 | 225 |
| Instacompute | 99.95 | 0.05 | 1.30103 | 14658 | 284 |

## 4.5 Problem 1: Maximum Availability –Fixed Budget

In order to replicate data a subset $S$ of data centers need to be chosen. We loose data when all the data centers in $S$ fail. Given a fixed budget $B$, how do we choose a subset $S$ of data centers, such that the cost of using these data centers does not exceed Budget $B$, and the probability of loosing data is minimized.

## 4.6 Assumptions

i) Every data center has sufficient amount of storage to store the entire data.
ii) All data is replicated in all data centers.
iii) The data centers have been filtered on various Qos parameters down to two dimensional QoS criteria, namely price and availability.
iv) Data centers are independent.
v) No. of service providers = 10.
vi) Available Budget <=50000 Dollars.
vii) Number of chunks <=40.
viii) Maximum availability needed is 99.999995% i.e., 1.57 seconds in a year.

## 4.7 Constraints

i) Cost of using chunk storage does not exceed Budget.
ii) Each service provider has sufficient space to store all the data.
iii) The service providers should have proper data security certifications.
iv) Response time as measured by other organizations for GET/PUT operations should be minimum.

## 4.8 Terminology

$D$ - set of data centers $d_1, d_2, d_3........d_n$

$B$ - the given budget

$S$ - subset $S$ of data centers where the data would be replicated.

$c(d)$ - cost of placing data in data center $d$

$c(S)$ - cost of using all data centers in $S$.

$p(d)$ - failure probability to recover data from data center $d$

$p(S)$ - failure probability to recover data from subset $S$ subset of data centers set $D$

## 4.9 Formulation

i) Cost of placing data in data centers in $S$

$$c(S) = \sum_{d \in S} c(d) \quad (1)$$

ii) Failure probability of all data centers in $S$

$$p(S) = \begin{cases} 1 & \text{if } S = 0 \\ \prod p(d) & \text{otherwise} \end{cases} \quad (2)$$

iii) Practicability:

A subset $S$ is practicable if the cost of $S$ is less than equal to Budget.

Cost of storing data should be less than the budget. That is,

$$c(S) \leq B \quad (3)$$

iv) Maximum availability fixed-budget problem: Predict the practicable subset $S^*$ of $D$ such availability of Cloud Storage is maximised.

v) Transform product into summation:

$l(d)$- a function which is equal to negative value of the logarithmic function on the failure probability of a data center $d$.

$$l(d) = -\log(p(d))$$

since $\log(x) + \log(y) = \log(xy)$

$$l(S) = -\log(p(S)) = \sum_{d \in S} l(d)$$

### 4.10 Transformation of problem into a knapsack problem

i) Given n service providers and a Budget $B$.

ii) Each service provider (data center) corresponds to an item.

iii) Cost $c(d)$ is the weight of the item.

iv) Negative logarithm $l(d)$ is the value of the item.

v) $c(d) > 0$ dollars and $l(d) > 0$

vi) Knapsack has a capacity equal to $B$ dollars.

vii) Goal: Fill knapsack so as to maximize total value.

viii) $V[i, b]$ = maximum profit subset of items $1..i$ with budget limit $b$.

ix) $V[i,b] = \begin{cases} 0 & \text{if } i = 0 \\ V[i-1,b] & \text{if } c(d)_i > b \\ \max\{V[i-1,b], (V[i-1, b-c(d)_i] + l(d)_i)\} & \text{otherwise} \end{cases} \quad (4)$

x) Failure Probability $= 1/pow(10, V[i,b])$

xi) Maximum Availability with budget $b$

$$b = 100 - (1/pow(10, V[i,b])) \qquad (5)$$

### 4.11 Service Providers Selection

If $V[i,b] \neq V[i-1,b]$ select the $i^{th}$ service provider with cost $c(d)_i$ and $b = b - c(d)_i$ and $i = i-1$ otherwise $i = i-1$.

### 4.12 Time Complexity

$O(nB)$ where $n$ is the number of service providers and $B$ is the budget.

### 4.13 Algorithm 1-Maximum Availability- Fixed Budget

```
Algorithm- 1
Input: W as budget ,n as number of cloud service providers, wt[] as cost of storage,
val[] as value of storage, V[] maximum value
Output: max value, item selection
for (w = 0 to W)
V[0,w] = 0
for (i = 0 to n)
V[i,0] = 0
for (i = 1 to n)
        for (w = 0 to W)
        if (wt[i] <= w)
        V[i,w] = max(val [i] + V[i-1,w- wt[i]],V[i-1,w])
        Else V[i,w] = V[i-1,w]
//select cloud service providers
i=n ,k=W
do
{
if  V[i, k] is not equal to V[i-1,k]
select item
i=i-1, k=k-wt[i]
else
i = i-1
}
while i>0
```

## 4.14 Problem 2: Maximum Expected Value-Fixed Budget

The previous maximum availability-fixed-budget problem demands a solution that minimizes the overall failure probability under a given budget. The data is replicated at more than one CSP.

Now, if there are *m* data chunks of any size and a given budget B, we want to replicate these *m* data chunks in such a way that the total cost of using these data centers does not exceed B, and the expected number of available data is maximized. We would like to find a practicable chunk assignment *S′* that maximizes the expected value of available chunks. The assignment is subject to various cost and performance considerations.

## 4.15 Assumptions

i) There is sufficient amount of storage space at each data centre if it is necessary to store all data chunks at one cloud storage.
ii) The data has been fragmented into chunks by the user according to privacy constraints.
iii) A data center also has two parameters – a *price per chunk availability* that we will be able to recover any data from it. Also the response time should be minimum.
iv) There are two replicas of each data chunk and any set two of data centers can be chosen for each chunk storage.

## 4.16 Terminology

*r* - number of replicas.
*m* - number of chunks.
*B* - available budget to store data.
*D* - set of data centers $d1, d2, d3........dn$.
*S* - subset *S* of data centers to replicate one chunk of data. $S_i \subseteq D$.
*b* - budget to store chunks first *(m−1)* chunks.
*c(d)* - cost of one chunk of data storage in data center *d*.
*p(d)* - failure probability to recover one chunk of data from data center *d* =100-availability.
*p(S)* - failure probability to recover data from subset *S*.
*c(S)* - cost of placing one chunk of data in each data center of *S*.
*E(S)* - the expected value of a data chunk in *S*.
*S′* - chunk assignment *S′* is a collection of all the data centers where all the *m* data chunks are replicated.

$c(S')$ - cost of all data centers selected for chunk assignment.

$E(S')$ - expected value of storing data among centers selected for chunk assignment $S'$.

$X$ - random variable that denotes expected value of chunk assignment under $S$.

### 4.17 Formulation

i) Cost of chunk assignment $S'$

$$c(S') = \sum_{i=1}^{m} c(S_i) \tag{6}$$

ii) Failure probability function: A relation between $p(d)$ and the expected value of the data chunks available at any given time under a chunk assignment $S'$ is derived.

iii) Expected value: Let $S$ be a set of data centers and $X$ be a random variable that denotes the total number of data chunks that are available if data chunks are replicated on all data centers in $S$. $X$ is referred to as *value* of a data chunk under $S$. The value of $X$ will be 0 if all the data centers where a data chunk is replicated are not available at any given time. The value of $X$ will be 1 if any of the data centers where a data chunk is replicated is available at any given time Thus the expected value of $X$, is as follows.

$$E(X) = 0 \times \prod_{d \in S} p(d) + 1 \times \left(1 - \prod_{d \in S} p(d)\right) \tag{7}$$

$$= 1 - \prod_{d \in S} p(d) \tag{8}$$

iv) $E(S)$ denotes the expected value of a data chunk when a chunk is replicated at any two data centers in $S$. Different chunks can be stored at different cloud service providers. Each cloud service provider can have distinct availability, the value of $E(S)$ for various chunks will also be distinct. It will be 0 if a chunk is not stored at all at any of the data centers.

$$E(S) = \begin{cases} 0 & \text{if } S = 0 \\ 1 - \prod_{d \in S} p(d) & \text{otherwise} \end{cases} \tag{9}$$

v) Expected value of all chunks of data under $S'$

$$E(S') = \sum_{i=1}^{m} E(S_i) \tag{10}$$

vi) Practicable: A chunk assignment $S'$ is practicable if $c(S') \leq B$.

## 4.18 Solution

A dynamic programming solution methodology is used to solve the maximum expected value of chunks - fixed-budget problem. The function $f(k,b)$ denoted the maximum expected value of available data chunks from the first $k$ chunks at cost $b$. The function $f(k,b)$ is computed using dynamic programming.

$$f(k,b) = \begin{cases} -\infty & \text{if } b < 0 \vee k < 0 \\ 0 & \text{if } b = 0 \wedge k = 0 \\ \max(f(k,b), f(k-1, b-c(S))) + E(S) & \text{otherwise} \end{cases} \quad (11)$$

## 4.19 Algorithm- 2 Maximum Expected Value –Fixed Budget

```
Algorithm-2 Maximum Expected Value-Fixed Budget
Input: n – number of service providers, r-replication factor, m-
number of chunks, W-budget, weightcs - cost of data chunks,
es- expected value of data chunk
Output: the maximum expected value of data that will be
available
float Z;
for b=0 to W
f[0,b]=0
for i=1 to m
f[i,0]=0
        for b=0 to M
                if(b>0) then
                        f[i,b]=f[i, b-1] // initialize
                        for each S is an element of C(n,r)
                        if (b-cS>=0)
                                Z=max( f[i,b],f[i-1,b-cS]+es)
                                if (f[i,b]<Z)
                                f[i,b]=Z
```

## 4.20 Time Complexity

The time complexity of Algorithm-2 is $O\left(mB \cdot \binom{n}{r}\right)$. It has been attempted to store data at cloud service providers with less response time and a fixed amount of budget, and also the replication factor for chunk assignment is taken as two. These QoS criteria and assumptions would reduce the complexity of the Algorithm-2 to $O(mB)$. The time mainly depends on number of data chunks and the Budget $B$ which are always fixed to a certain value. As a result the algorithm runs at a very high speed even on a desktop computer.

# Chapter 5

# EXPERIMENTAL SETUP AND RESULTS

The experiments were conducted on an Intel® Core(TM) i3 CPU 540 @ 3.07GHz. The desktop has 2 GB memory. C++ code of 8.3 KB with 415 lines was compiled with g++ compiler on ubuntu 13.04 to find out maximum availability within a given budget and selecting the multiple cloud service providers who can provide this availability and also to verify the time elapsed in finding out the solution. The number of data centers taken for the experiment are 10 and the budget is up to 50000 Dollars to store 52 TB of data as this is sufficient to achieve 7 nines availability.

The size of data for cloud storage and budget for storage are provided as input data. The outputs are maximum value, maximum availability, cloud service providers selected within given budget and the time elapsed in milliseconds.

To find out the maximum expected value of available data chunks within a given budget it is assumed that we have a maximum number of 40 data chunks to store in 10 possible data centers as identified by minimum response time and security certifications, and each data chunk is replicated at a maximum of two cloud service providers.

The amount of data to be stored, budget for storage as well as the number of chunks to store are provided as input data. The data regarding Cloud service providers such as security certifications, cost of storage, availability has been collected from their respective websites. The data regarding response time has been collected from Cedexis.com which performs various intensive tests to predict the response time of various service providers at different regions of the world.

Maximum Availability-Fixed Budget: The dynamic programming Algorithm-1 was implemented for solving the maximum availability fixed budget problem. The results obtained for storing 50 TB of data are illustrated in Table 5.1.

*Table 5.1: Budget, Availability and Time Elapsed*

| Budget | Max Value | Failure Probability | Availability | CSP Selected | Time in milliseconds |
|--------|-----------|---------------------|--------------|--------------|----------------------|
| 9950   | 1         | 0.1                 | 99.9         | 6            | 466.60               |
| 11350  | 1.30103   | 0.05                | 99.95        | 8            | 544.33               |
| 11700  | 2         | 0.01                | 99.99        | 3            | 565.67               |
| 21300  | 2.30103   | 0.005               | 99.995       | 8,6          | 1078.23              |
| 21650  | 3         | 0.001               | 99.999       | 6,3          | 1114.29              |
| 23050  | 3.30103   | 0.0005              | 99.9995      | 8,3          | 1177.29              |
| 25100  | 4         | 0.0001              | 99.9999      | 3,1          | 1411.76              |
| 33000  | 4.30103   | 0.00005             | 99.99995     | 8,6,3        | 1844.07              |
| 36400  | 5.30103   | 0.00001             | 99.99999     | 8,3,1        | 2113.30              |

An availability of 99.9 % to 99.99 % can be achieved by storing data at only one storage provider. An availability of 99.995% is easily available by duplicating the data at two most cheapest storage providers as shown in Figure 5.1 and an availability of 99.9999 % is achieved by duplicating the data at the best storage providers with only a slight increase in the budget. To achieve availability greater than 99.9999%, a substantial increase in the budget is required. But this is only required for a very few applications only.

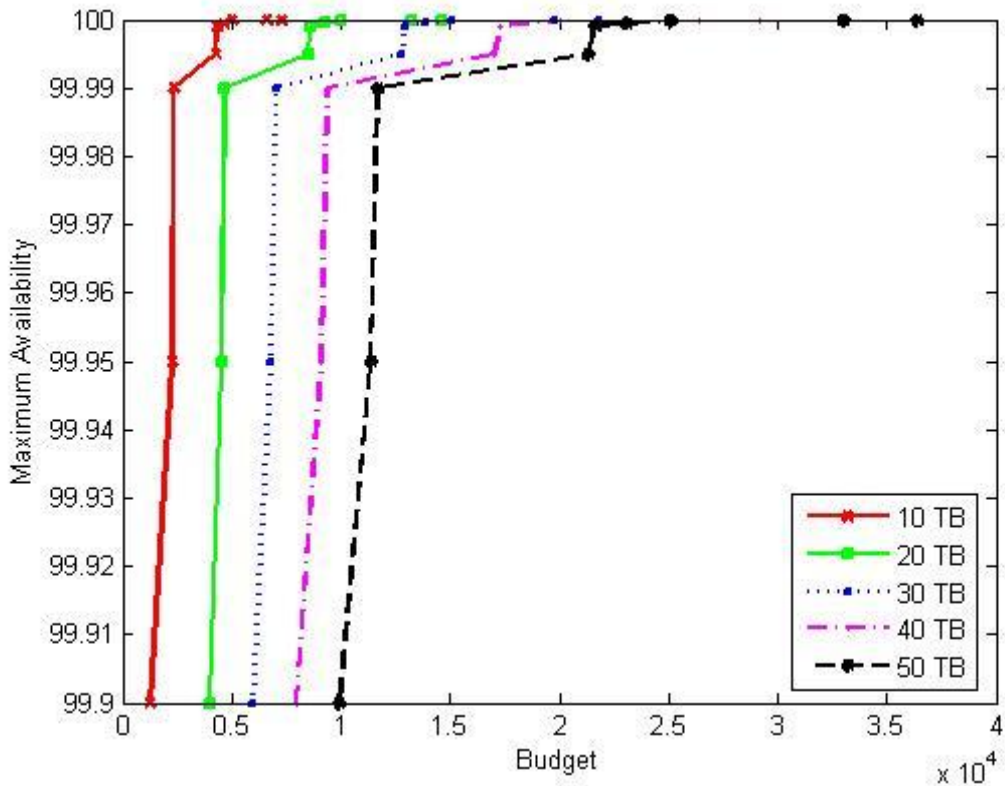

*Figure 5.1: Budget versus Maximum Availability*

The failure probability of the cloud storage providers is also plotted against the given budget on logarithmic scale as shown in Figure 5.2 . As the budget increases the failure probability on logarithmic scale (Maximum Value) also increases due to two reasons. Firstly, at higher budget we can avail services of a better service provider. Secondly, when we increase the budget further, we can get services of more than one, two or three service providers.

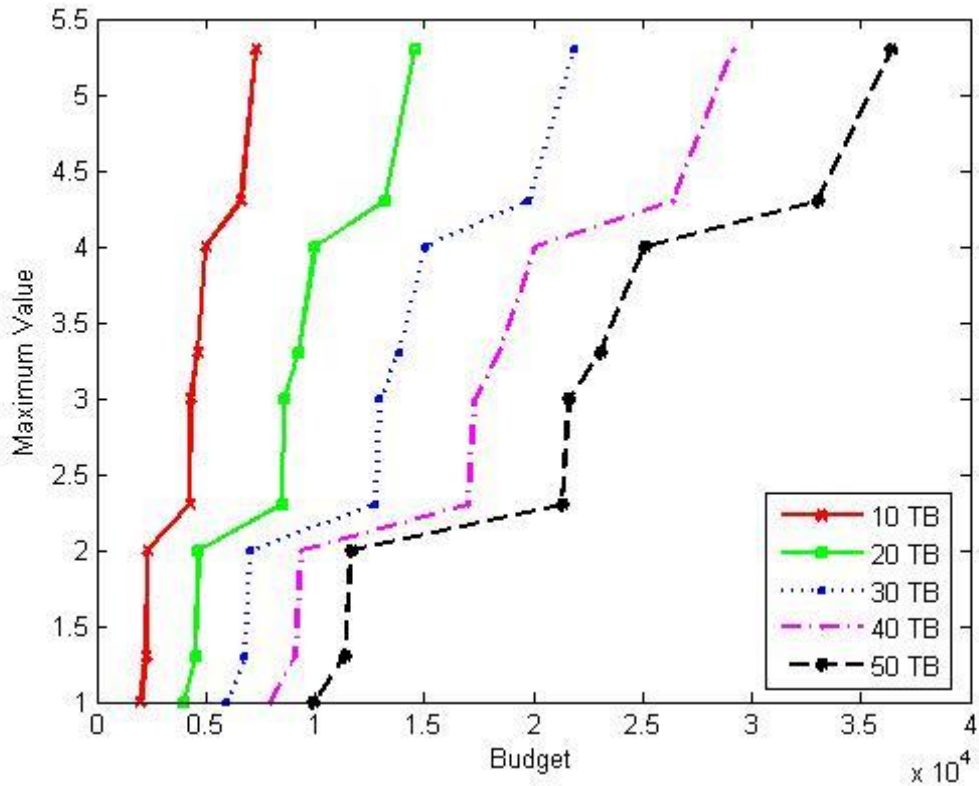

*Figure 5.2: Budget versus Maximum Value on logarithmic scale*

The Figure 5.3 illustrates the execution time of Algorithm-1. The execution time increases as the budget increases. This is due to the fact that higher availability is provided at higher budget. The graph is not a straight line but has several steep steps also. This is due to the reason, that it selects different providers at a very small increase in cost, but when the cost increases by a large amount, the number of service providers also increase. The algorithm in that case takes more time in selecting the Cloud service providers. The execution time is dependent only on the budget and the number of cloud service providers and not the sequence of the service provider in which it is listed as the algorithm selects the service providers only after computing the total value for all the service providers.

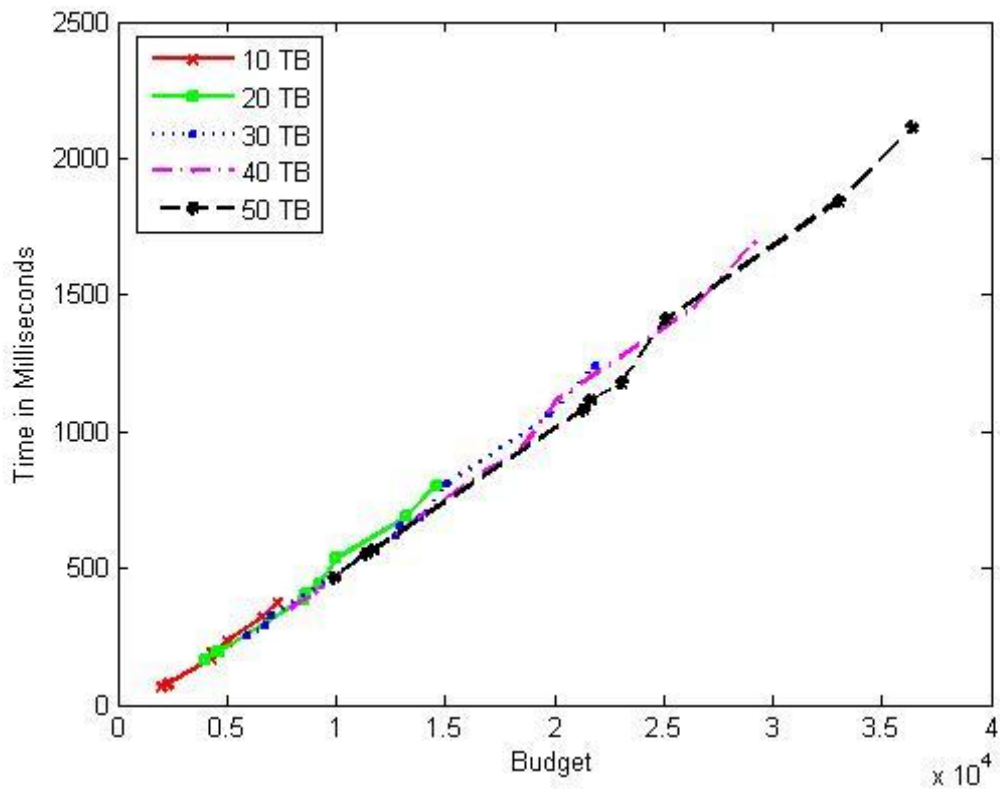

*Figure 5.3: Budget versus Execution time in milliseconds*

Maximum Expected Value-Fixed Budget: Table 5.2 shows Budget versus Expected value of each data chunk for storing 40 TB of data. As the Budget increases, the expected value of each data chunk when each chunk is replicated at two service providers increases. This is due to the fact that better services are provided at a higher cost. And the algorithm searched the best cloud service providers. But, it is not always true that by spending more amount one can get more expected value. Even then, it may be required to taken services of that provider, if it is desired to store different chunks at different providers due to privacy constraints or response time benefits. This is shown graphically in Figure 5.4.

*Table 5.2: Budget versus Expected Value*

|  | Expected Value (After Replication) |
|---|---|
| 17040 | 0.995 |
| 17320 | 0.999 |
| 18440 | 0.9995 |
| 18680 | 0.999 |
| 19800 | 0.9995 |
| 20080 | 0.9999 |

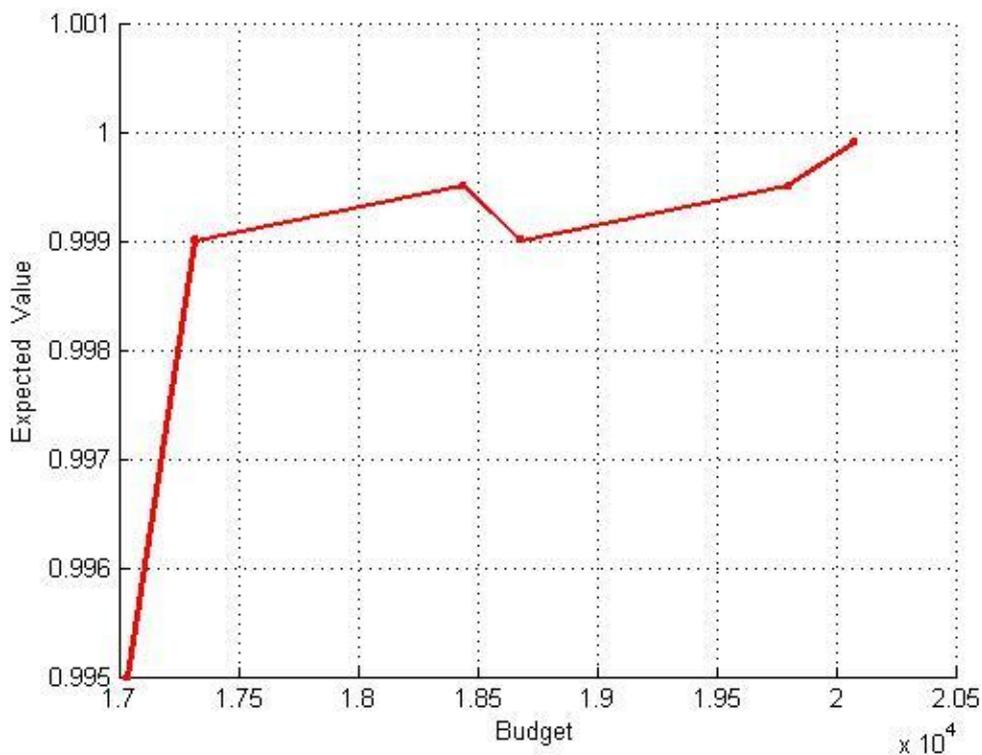

*Figure 5.4: Budget Versus Expected Value of each data chunk*

Table 5.3 shows number of chunks versus time in milliseconds to find out the maximum expected value of all the chunks. As the number of chunks increase, the time taken to find the number of valid data chunks increase. This is due to the fact that for each chunk one more iteration is required. The time increases with budget also because for each chunk more number of iterations are required. This is shown graphically in Figure 5.5.

*Table 5.3: Number of chunks versus time in milliseconds*

| Chunks | Time in milliseconds |
|---|---|
| 1 | 11.13 |
| 2 | 19.49 |
| 3 | 19.55 |
| 10 | 73.89 |
| 20 | 164.79 |
| 30 | 240.97 |
| 40 | 331.87 |

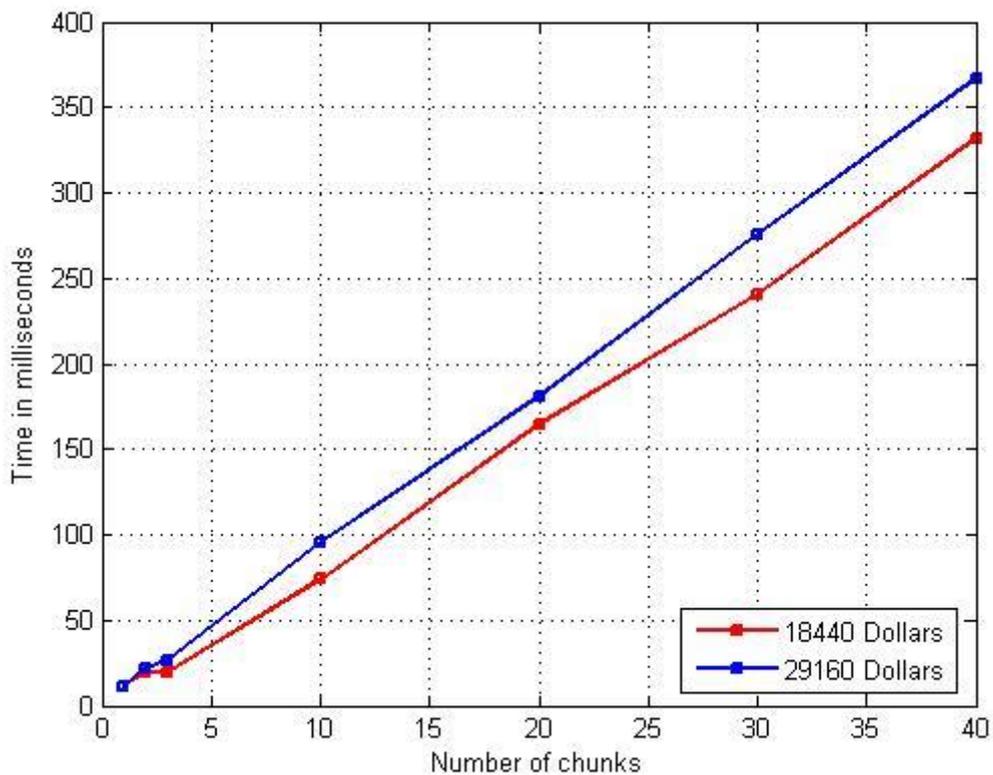

*Figure 5.5: Number of chunks versus Time in milliseconds*

# Chapter 6

# CONCLUSION

A new secured, cost effective, highly available multi-cloud architecture for enabling privacy-preserving outsourced storage of data has been introduced. The solution seeks to provide each customer with a better cloud data storage decision, taking into consideration the user budget as well as providing him with the best quality of service (security, response time and availability of data) offered by available cloud service providers.

Fragmentation of data into chunks which preserve privacy are used to decompose data which makes the data invaluable even if an intruder gets access to this data in this multi-cloud architecture. A definition of privacy based on hiding sets of attribute values have been demonstrated, and how the decomposition techniques help in achieving privacy is also shown.

Given the increasing instances of cloud DbaaS, as well as the increasing prominence of privacy concerns as well as regulations, it is expected that the architecture used will prove useful both in ensuring compliance with laws and in reducing the risk of privacy breaches.

This paper also addresses the issues of selecting multiple cloud providers. The motivation of having multiple cloud providers has been described, and a mathematical model for evaluating the quality of those algorithms that select service providers has been formally defined. This model addresses both the object functions and cost measurements, in which optimization problems on their trade-off can be defined. Based on the model, two algorithms for selecting service providers with a given budget have been derived. One algorithm maximizes data survival probability, and the other maximize the expected value of the available data blocks. Experiments have been conducted to demonstrate that the proposed algorithms are efficient enough to find optimal solutions in reasonable amount of time.

From the experiments, it is observed that replication is extremely effective in improving data availability. Using multiple data providers that have much high failure probability than the leading provider is sufficient to guarantee high availability.

By selecting data centers from different providers to host our cloud servers, the hazards related to business continuity can be removed altogether, as well as issues concerning electricity suppliers, networking providers and other "data center" issues can be resolved, since all the cloud providers operate independently.

A multi-cloud strategy also reduces other risks associated with having a single provider. It also reduces the one time cost of switching storage providers in exchange for additional operational overhead.

Through careful selection of cloud storage vendors, it is possible to tolerate outages and mitigate vendor lock-in with reasonable over-head cost within a given budget.

**LIMITATIONS**

1) The run time of the algorithm is strongly dependent on the budget, so we cannot guarantee bounds on the runtime.

2) This work does not provide transparent access mechanism to recover data among multiple cloud providers.

3) Does not include the effects due to replication.

4) Does not include extra communication cost for synchronizing replicas.

5) Does not include the cost of switching cloud service providers.

# APPENDIX - 1

**Failure probability and the cost of storage of popular CSPs**

| SN | CSP | Availability | Data Storage | | | Data Transfer Out | | | | PUT, COPY, POST, or LIST Requests per 1000 requests | GET and all other Request per 10000 requests |
|---|---|---|---|---|---|---|---|---|---|---|---|
| | | | First 1 TB /month $ per GB | Next 49 TB /month $ per GB | Next 450 TB /month $ per GB | First 1 GB /month $ per GB | Upto 10 TB /month $ per GB | Next 45 TB/ month $ per GB | Next 100 TB /month $ per GB | | |
| 1 | Amazon | 99.99 | 0.095 | 0.080 | 0.070 | 0.000 | 0.190 | 0.150 | 0.130 | 0.005 | 0.004 |

| SN | CSP | Availability | Data Storage | | | | Data Transfer Out | | | PUT, COPY, POST, or LIST Requests per 1000 requests | GET and all other Request per 10000 requests |
|---|---|---|---|---|---|---|---|---|---|---|---|
| | | | First 1 TB /month $ per GB | Next 9 TB /month $ per GB | Next 90 TB /month $ per GB | Next 400 TB /month $ per GB | First 1 TB /month $ per GB | Next 9 TB /month $ per GB | Next 90 TB/ month $ per GB | | |
| 2 | Google App Engine | 99.90 | 0.085 | 0.076 | .067 | 0.063 | 0.21 | 0.18 | 0.15 | 0.01 | 0.01 |

| SN | CSP | Availability | Data Storage | | | | Data Transfer Out | | | | PUT, COPY, POST, or LIST Requests per 1000 requests | GET and all other Request per 10000 requests |
|---|---|---|---|---|---|---|---|---|---|---|---|---|
| | | | 1-10 GB /month $ per GB | 10 GB – 1 TB /month $ per GB | 1 TB- 50 TB /month $ per GB | 50 TB- 500 TB /month $ per GB | First 1 GB /month $ per GB | GB - 1 TB/ month $ per GB | 1 TB- 10TB/ month $ per GB | 10TB- 50 TB /month $ per GB | | |
| 3 | GoGrid | 99.99 | 0.00 | 0.12 | 0.11 | 0.10 | 0.000 | 0.12 | 0.11 | 0.10 | free | free |

| SN | CSP | Availability | Data Storage | | | | Data Transfer Out | | | | PUT, COPY, POST, or LIST Requests per 1000 requests | GET and all other Request per 10000 requests |
|---|---|---|---|---|---|---|---|---|---|---|---|---|
| | | | First 1 TB /month $ per GB | Next 49 TB /month $ per GB | Next 150 TB /month $ per GB | Next 300 TB /month $ per GB | First 1 GB /month $ per GB | GB - 1 TB/ month $ per GB | 1 TB- 10TB/ month $ per GB | 10TB- 50 TB /month $ per GB | | |
| 4 | Rackspace | 99.9 | 0.10 | 0.09 | 0.085 | 0.080 | 0.12 | 0.12 | 0.12 | 0.12 | free | free |

| SN | CSP | Availability | Data Storage | | | | Data Transfer Out | | | | PUT, COPY, POST, or LIST Requests per 1000 requests | GET and all other Request per 10000 requests |
|---|---|---|---|---|---|---|---|---|---|---|---|---|
| | | | First 1 TB /month $ per GB | Next 49 TB /month $ per GB | Next 150 TB /month $ per GB | Next 300 TB /month $ per GB | First 1 GB /month $ per GB | 2 GB - 1 TB/ month $ per GB | 1 TB- 10TB/ month $ per GB | 10TB- 50 TB /month $ per GB | | |
| 5 | HP Cloud | 99.95 | 0.10 | 0.10 | 0.10 | 0.10 | 0.16 | 0.11 | 0.09 | 0.07 | 0.01 | 0.01 |

| SN | CSP | Availability | Data Storage | | | | Data Transfer Out | | | | PUT, COPY, POST, or LIST Requests per 1000 requests | GET and all other Request per 10000 requests |
|---|---|---|---|---|---|---|---|---|---|---|---|---|
| | | | First 1 TB /month $ per GB | Next 49 TB /month $ per GB | Next 450 TB /month $ per GB | Next 500 TB /month $ per GB | First 1 GB /month $ per GB | 2 GB - 1 TB/ month $ per GB | 1 TB- 10TB/ month $ per GB | 10TB- 50 TB /month $ per GB | | |
| 6 | Microsoft Windows Azure | 99.9 | 0.095 | 0.08 | 0.07 | 0.065 | 0.12 | 0.08 | 0.06 | 0.04 | 0.01 | 0.01 |

| SN | CSP | Availability | Data Storage | | | | Data Transfer Out | | | | PUT, COPY, POST, or LIST Requests per 1000 requests | GET and all other Request per 10000 requests |
|---|---|---|---|---|---|---|---|---|---|---|---|---|
| | | | First 1 TB /month $ per GB | Next 49 TB /month $ per GB | Next 450 TB /month $ per GB | Next 500 TB /month $ per GB | First 1 GB /month $ per GB | 2 GB - 1 TB/ month $ per GB | 1 TB-10TB/ month $ per GB | 10TB-50 TB /month $ per GB | | |
| 7 | Savvis Cloud | 99.95 | 0.095 | 0.095 | 0.095 | 0.095 | 0.12 | 0.12 | 0.12 | 0.12 | 0.01 | 0.01 |

| SN | CSP | Availability | Data Storage | | | | Data Transfer Out | | | | PUT, COPY, POST, or LIST Requests per 1000 requests | GET and all other Request per 10000 requests |
|---|---|---|---|---|---|---|---|---|---|---|---|---|
| | | | 0-50 TB /month $ per GB | Next 50 TB /month $ per GB | Next 150 TB /month $ per GB | Next 300 TB /month $ per GB | First 1 GB /month $ per GB | 2 GB - 1 TB/ month $ per GB | 1 TB-10TB/ month $ per GB | 10TB-50 TB /month $ per GB | | |
| 8 | Inter nap Agile FILES | 99.95 | 0.10 | 0.09 | 0.09 | 0.09 | 0.10 | 0.10 | 0.10 | 0.10 | 0.01 | 0.01 |

| SN | CSP | Availability | Data Storage | | | | Data Transfer Out | | | | PUT, COPY, POST, or LIST Requests per 1000 requests | GET and all other Request per 10000 requests |
|---|---|---|---|---|---|---|---|---|---|---|---|---|
| | | | 0-50 TB /month $ per GB | Next 50 TB /month $ per GB | Next 150 TB /month $ per GB | Next 300 TB /month $ per GB | First 1 GB /month $ per GB | 2 GB - 1 TB/ month $ per GB | 1 TB-10TB/ month $ per GB | 10TB-50 TB /month $ per GB | | |
| 9 | SoftLayer | 99.9 | 0.10 | 0.10 | 0.10 | 0.10 | 0.10 | 0.10 | 0.10 | 0.10 | 0.01 | 0.01 |

| SN | CSP | Availability | Data Storage | | | | Data Transfer Out | | | | PUT, COPY, POST, or LIST Requests per 1000 requests | GET and all other Request per 10000 requests |
|---|---|---|---|---|---|---|---|---|---|---|---|---|
| | | | 0-50 TB /month $ per GB | Next 50 TB /month $ per GB | Next 150 TB /month $ per GB | Next 300 TB /month $ per GB | First 1 GB /month $ per GB | 2 GB - 1 TB/ month $ per GB | 1 TB- 10TB/ month $ per GB | 10TB- 50 TB /month $ per GB | | |
| 10 | Instacompute | 99.95 | 0.115 | 0.115 | 0.115 | 0.115 | 0.135 | 0.135 | 0.135 | 0.135 | 0.01 | 0.01 |

**Availability, Failure Probability and –log(Availability)**

| Availability | Failure Probability | -log(Availability) |
|---|---|---|
| 99.9 | 0.100000 | 1.000000 |
| 99.95 | 0.050000 | 1.301030 |
| 99.99 | 0.010000 | 2.000000 |
| 99.995 | 0.005000 | 2.301030 |
| 99.999 | 0.001000 | 3.000000 |
| 99.9995 | 0.000500 | 3.301030 |
| 99.9999 | 0.000100 | 4.000000 |
| 99.99995 | 0.000050 | 4.301030 |
| 99.99999 | 0.000010 | 5.000000 |
| 99.999995 | 0.000005 | 5.301030 |